\documentclass[twocolumn]{article}
\usepackage{graphicx}
\usepackage[T1]{fontenc}
\usepackage{mathptmx}
\usepackage[scaled=.92]{helvet}
\usepackage{courier}

\usepackage{authblk}

\newcommand*{\correspondence}[2]{%
  \gdef\AESCorrAuthor{#1}%
  \gdef\AESCorrEmail{#2}%
}%
\newcommand*{\lastnames}[1]{\gdef\AESLastnames{#1}}%
\newcommand*{\shorttitle}[1]{\gdef\AESShortTitle{#1}}%

\graphicspath{{./}{figures/}}

\usepackage{multicol}
\usepackage[export]{adjustbox}

\usepackage[utf8]{inputenc}

\usepackage{microtype}

\usepackage[numbers,square]{natbib}

\usepackage{booktabs}
\usepackage{color}
\usepackage{url}

\usepackage{textcomp}
\usepackage{mathptmx}
\usepackage{amsmath,amssymb}
\usepackage{gensymb}
\usepackage{verbatim}
\usepackage{balance}

\usepackage{url}
\usepackage{xspace}

\usepackage{paralist}
	\renewenvironment{itemize}[1]{\begin{compactitem}#1}{\end{compactitem}}
	\renewenvironment{enumerate}[1]{\begin{compactenum}#1}{\end{compactenum}}

\usepackage[compact]{titlesec}
    \titlespacing{\section}{0pt}{2ex}{1ex}
    \titlespacing{\subsection}{0pt}{1ex}{0ex}
    \titlespacing{\subsubsection}{0pt}{0.5ex}{0ex}
    
\renewcommand{\vec}[1]{\ensuremath{\mathbf{#1}}\xspace}

\newcommand{\uvec}[1]{\ensuremath{\vec{\hat{#1}}}\xspace}

\newcommand{\rE}[0]{\ensuremath{r_{E}}\xspace}
\newcommand{\rEvec}[0]{\vec{r_E}}

\newcommand{\rV}[0]{\ensuremath{r_{V}}\xspace}
\newcommand{\rVvec}[0]{\vec{r_{V}}}

\newcommand{\figureref}[1]{\figurename\,\ref{#1}}

\newcommand{\TODO}[1]{}
\title{Optimized Decoders for Mixed-Order Ambisonics\footnote{This paper has been accepted for presentation at the 150th Audio Engineering Society Convention, May 25-28, 2021, (virtual).}}

\author[1]{Aaron Heller}
\author[2]{Eric Benjamin}
\author[3]{Fernando Lopez-Lezcano}

\affil[1]{Artificial Intelligence Center, SRI International, Menlo Park, CA}
\affil[2]{Surround Research, Pacifica, CA}
\affil[3]{Center for Computer Research in Music and Acoustics (CCRMA), Stanford University, Stanford, CA}

\correspondence{Aaron Heller}{aaron.heller@sri.com}

\lastnames{Heller, Benjamin, and Lopez-Lezcano}

\shorttitle{Optimized Ambisonic Decoders}

\date{May 23, 2021}

\begin{document}
\maketitle % MANDATORY!

\footnotetext[1] {\url{aaron.heller@sri.com}}
\footnotetext[2] {\url{ericmbenj@gmail.com}}
\footnotetext[3] {\url{nando@ccrma.stanford.edu}}

\begin{abstract}
  %\small
  %\narrower
  In this paper we discuss the motivation, design, and analysis of
  ambisonic decoders for systems where the vertical order is less than
  the horizontal order, known as mixed-order Ambisonic systems.  This
  can be due to the use of microphone arrays that emphasize horizontal
  spatial resolution or speaker arrays that provide sparser coverage
  vertically. First, we review Ambisonic reproduction criteria, as
  defined by Gerzon, and summarize recent results on the relative
  perceptual importance of the various criteria.  Then we show that
  using full-order decoders with mixed-order program material results
  in poorer performance than with a properly designed mixed-order
  decoder. We then introduce a new implementation of a decoder
  optimizer that draws upon techniques from machine learning for quick
  and robust convergence, discuss the construction of the objective
  function, and apply it to the problem of designing two-band decoders
  for mixed-order signal sets and non-uniform loudspeaker
  layouts. Results of informal listening tests are summarized and
  future directions discussed.
\end{abstract}

\section{Introduction}

There is a renewed interest in decoders for mixed-order Ambisonics due
to the availability of mixed-order microphones and the current
COVID-19 restrictions placing an emphasis on loudspeaker arrays that
can be deployed in domestic settings, where it is relatively easy to
deploy a third-order horizontal array comprising eight loudspeakers.
However, installing more than a few elevated speakers is difficult and
placing speakers significantly below the listener is nearly
impossible. While mixed-order operation is frequently cited as an
advantage of Ambisonics, little has been written about creating or
analyzing the performance of decoders specifically for mixed-order
signal sets or highly non-uniform loudspeaker arrays. We also
introduce a new implementation of the Ambisonic Decoder Toolbox (ADT) in
Python/NumPy, which includes a fast and robust non-linear optimizer
and a new design procedure for dual-band decoders where we first
optimize the high-frequency performance of the decoder and then
optimize the low-frequency performance to match the high-frequency  
\cite{www:adt}.

\section{Ambisonics}

%% N: do we really need an introduction to ambisonics?
Ambisonics is an extensible, hierarchical system for representing
sound fields. It defines how something should sound as opposed to
specifying the signals going to particular speakers. Sound fields can
be recorded using an Ambisonic microphone or created using an
Ambisonic panner to position a sound in full 3D space. It is an
isotropic representation of the sound field that can be rotated in the
renderer making it attractive for virtual and augmented reality
applications.

An Ambisonic signal set is a representation of the sound field as the
time-varying coefficients of a spherical harmonic series. The spatial
accuracy increases with the number of harmonics being used. A
first-order Ambisonic signal set is four channels wide, third-order is sixteen
channels, fifth order is 36, and so forth. Each increase in Ambisonic
order adds spherical harmonics to the signal set and increases the
spatial accuracy of the representation of the sound field. We use a
shorthand notation to specify the signal set. For example 3H2V means
third-order horizontal, second-order vertical, with the set of
spherical harmonics according to the HV convention
\cite{Travis:2009vl}.

Once an Ambisonic signal set has been captured or generated,
appropriate speaker feeds are produced by a decoder. Designing an
optimal decoder, specifically the low- and high-frequency matrices,
for a given signal set and loudspeaker array is the central topic of
this paper. Other aspects of decoder design have been covered in
earlier papers by the present authors \cite{Heller:2018vh}.

\subsection{Mixed-Order Ambisonics}
A physical encoder (an Ambisonic microphone) needs to have enough
capsules covering the sphere to accurately sample the spherical
harmonics of the order it is intended to capture. Conversely, a
speaker array needs to have enough loudspeakers covering the sphere to
excite the spherical harmonics for the maximum order it is intended to
reproduce. That is not always the case, leading to arrays with 
different densities of transducers in different directions. The
consequence is that the order that can be encoded or decoded will
change according to the direction.

For example, nine years ago, one of the present authors published the
design for a second-order ambisonic microphone \cite{Benjamin:2012vp}.
There have been four proprietary \cite{web:octomic, web:voyage,
  www:Brahma8,www:Reynolds8} and one free and open-source
implementation \cite{LopezLezcano:2019hn} of this design. A compromise
made was to use only eight capsules. This simplifies calibration and
allows the use of widely-available eight-channel recorders.

While commonly referred to as a second-order microphone, only eight of
the nine spherical harmonic components needed for the second-order
signal set can be derived from the capsule signals. The missing
spherical harmonic is degree 2 and order 0, which is called ``R'' in the
Furse-Malham convention. R is a ``zonal'' harmonic and varies only
with elevation. Eliminating this component coarsens the description of
the sound field at elevations other than horizontal, making it a 2HV1 
mixed-order encoder.  As we shall see, decoding this signal set with a decoder designed for
full second order is suboptimal.

%% << N: relevant? where do we ask this? >> The question, then, is;
%% how significant is the impairment, how do we design optimal
%% decoders for it, and, more generally, is the microphone useful for
%% full-sphere audio recording?

Small speaker arrays with a limited number of speakers in the vertical
direction are another case in which the array does not have uniform
density of speakers and cannot excite the spherical harmonics
in all directions equally. Physical restrictions in the placement of speakers
can also dictate that an array might not be capable of rendering the same
order in both the horizontal and vertical directions. Such an array
will need a mixed-order decoder.

% We need to be able to create encoders and decoders that can
% correctly encode and decode Ambisonics signal sets of
% mixed-order. In this paper we will focus on the design of decoders
% for mixed-order systems.

%% has this even been generalized? mixed-order is always thought of as
%% differences between horizontal and vertical orders, but that is
%% arbitrary, right? It could be general in the sense that the order
%% depends on the density of transducers globaly - in any direction
%% (which would be the same as your graphs of "relative order" that we
%% use to evaluate decoders. There could be an array that is "front
%% rich" in number of speakers and would need a mixed-order decoder,
%% only the decoder would have higher otder towards the front (hmmm,
%% that would be 5.1/7.1, right???)

\section{Ambisonic Decoders}
The task of the decoder is to create the best perceptual impression
possible that the sound field is being reproduced accurately, given
the available resources.  In practical terms, the following criteria are
necessary\cite{Gerzon92:Metatheory}:
\begin{enumerate}
\item Constant amplitude gain for all source directions
\item Constant energy gain for all source directions
\item At low-frequencies, correct reproduced wavefront direction and
  velocity (Gerzon's velocity-model localization vector, $\rVvec$)
\item At high-frequencies, maximum concentration of energy in the
  source direction (Gerzon's energy-model localization vector,
  $\rEvec$)
\item Matching high- and low-frequency perceived directions
  ($\uvec{\rE} = \uvec{\rV}$)
\end{enumerate} %
Recent work shows that (4) is the most important \cite{Frank:2013ud}; it
is also the most difficult to get right. After that, (2) and (5) are
important, as it is thought that we use a majority voting system to
resolve conflicting directional cues
\cite{Gerzon92:Metatheory}. Decoders that ignore (5) can be fatiguing due to 
conflicting perceptual cues \cite{Scharine:2009}. Note that
to satisfy all of these criteria we must use decoders that have
different gain matrices for high and low frequencies, so-called
``two-band'' or ``Vienna'' decoders \cite{Gerzon92:DecodersHDTV}.

The ADT includes a full-featured decoder engine written in the FAUST
DSP specification language \cite{Smith:aspf} that implements dual-band
decoding, near-field correction, and level and time-of-arrival
compensation. The ADT incorporates several design techniques that
produce decoders that perform well according to these criteria for
partial-coverage loudspeaker arrays, such as domes and stacked rings,
but assumes that within those limits the speakers are (more or less)
uniformly distributed. It also assumes that the decoders produced by
these techniques are optimal for mixed-order signal sets.

\subsection{Mixed-Order Decoders}

\begin{figure*}[ht!]
\centering
\begin{tabular}{c c}
    \includegraphics[width=0.45\textwidth]{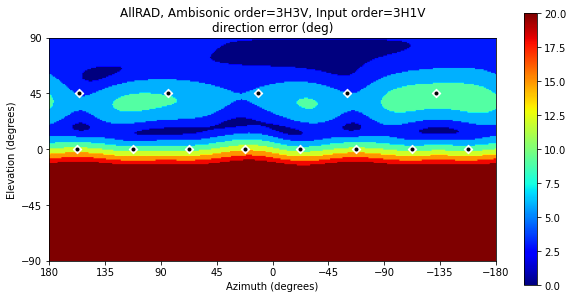} &
    \includegraphics[width=0.45\textwidth]{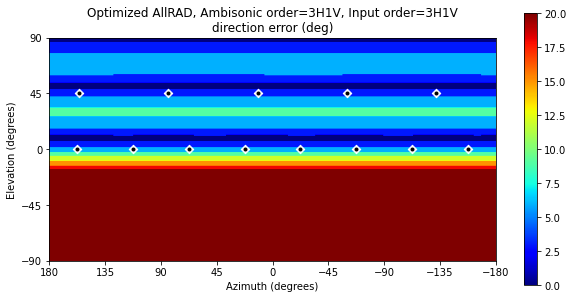} \\
    (a) A 3H1V signal rendered by a 3H3V decoder. & (b) 3H1V signal rendered by a 3H1V decoder.\\
    \includegraphics[width=0.45\textwidth]{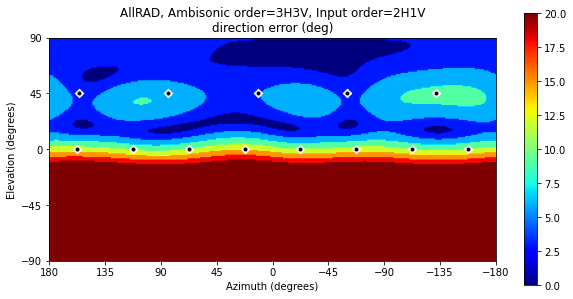} &
    \includegraphics[width=0.45\textwidth]{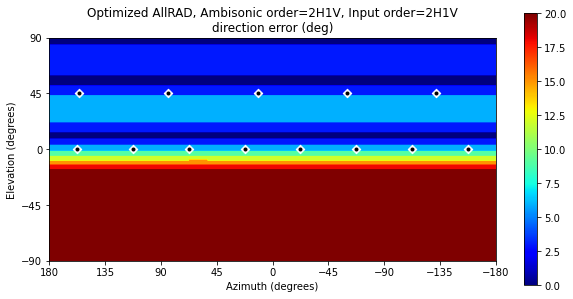} \\
    (c) A 2H1V signal rendered by a 3H3V decoder. & (d) 2H1V signal rendered by a 2H1V decoder.\\
\end{tabular}
\caption{The effect on directional accuracy using a generic 3H3V
  decoder vs. decoders designed for the specific signal set in use. 
  The speaker array has 8 horizontal and 5 height speakers. The
  dots show the locations of the loudspeakers.}
\label{fig:mixed}
\end{figure*}

Many diffusion systems simply use full-order decoders for mixed-order signal
sets, leaving the missing channels unconnected, such as ``R'' in the case of
the eight-capsule microphone described above.  In Ambisonics, omitting
channels from a signal set and leaving those channels unconnected and
silent in a full-order signal set are two different things.  In the
former case, the decoder assumes a point source where the omitted
components are not known. In the latter case, the decoder assumes that
those components are known and that the spatial distribution of the
sources is such that the silent components are exactly zero. The
latter case would be extremely rare in real acoustic scenes.

This was investigated and we found that in every case examined, a
decoder specifically designed for the mixed-order signal set
outperformed a full-order decoder by the criteria listed at the beginning of this section.

As an example, \figureref{fig:mixed} shows the directional error of
3H1V and 2H1V signal sets being decoded by a 3H3V decoder (a and c) vs.
a mixed-order decoder designed for the specific signal set (b and d).
The speaker array is a small 8+5 array. A mixed-order decoder performs
much better in terms of accuracy of the rendered directions.

%% these four figures were lifted from the slides for the talk, I do not have
%% the source that was used to create them
%%
%% NOTE: does not seem fair as these are comparing a plain 3H3V AllRAD with an
%% optimized decoder of the proper order. It would be better to see how a 3H3V
%% optimized decoder compares to, say, a 3H1V optimized decoder.

\section{Designing Decoders}
For regular speaker arrays (2D polygons, 3D polyhedra, t-designs) the design of a correct decoder is a straightforward task: %
\begin{itemize}
\item Build the speaker encoding matrix, \vec{K}, by sampling the spherical harmonics at the speaker directions.
\item Use the pseudoinverse to find the basic decoding matrix, \vec{M}.
\item Modify the per-order gains of \vec{M} to maximize the magnitude of \rEvec.
\end{itemize}
For this type of array, Gerzon proved that \rEvec will point in same direction as \rVvec \cite{Gerzon92:Metatheory}.

\subsection{Partial-Coverage Arrays}%
In most cases, except perhaps for 2D arrays, it is hard to deploy
truly regular speaker arrays. Physical constraints limit the placement
of speakers, 3D arrays need speakers above and below the listener, and
smaller arrays usually have less density of speakers in the vertical
direction. Most practical 3D arrays are domes or stacked rings with no
speakers below the listener area. The ADT implements several design
techniques that produce decoders that perform well for these
partial-coverage loudspeaker arrays:%
\begin{itemize}
\item Use an inversion technique suited to ill-conditioned matrices.
\item Derive a new set of basis functions for which inversion is well
  behaved, EPAD \cite{Zotter:2012ev}.
\item Invert a well-behaved full-sphere virtual speaker array, map to
  a real array, AllRAD \cite{Zotter:2012uk}.
\end{itemize}%
Currently, the AllRAD method is able to cope well with partial-coverage
arrays and is our  ``go-to'' technique to design decoders for them.

In general, these techniques trade off localization accuracy for
uniform loudness. Typically, \rEvec and \rVvec will not point in the
same direction and localization quality degrades in areas of low
density of speakers. These tradeoffs are determined by the particular
technique in use and are not directly under a user's control. Many,
starting with Gerzon \cite{Gerzon92:DecodersHDTV}, have turned to
numerical optimization techniques to enable more direct control over
these tradeoffs. The literature is full of descriptions of
implementations, some taking days to find a solution, but, as far as
we know, the implementation described in the next section is the first
one released publicly that is capable of producing two-band
decoders.

\section{Optimizing Decoders}
To address these shortcomings, we implemented a decoder-matrix
optimizer that directly implements the above Ambisonic decoder
evaluation criteria in its objective function.  Each of the criteria
is evaluated at 5200 points of a spherical design
\cite{wiki:spherical-design}, spatially weighted according to the
loudspeaker coverage area, then summed to produce the value of the
objective function. Because some of the criteria are non-linear, we
employ a constrained, quasi-Newton method, L-BFGS \cite{wiki:lbfgs},
which is available in SciPy's optimization module. Additionally, we
employ the JAX library for automatic differentiation of Python/NumPy
code to perform the gradient calculation needed by
L-BFGS\cite{www:jax}. This provides a large speed-up over
finite-difference techniques and will use a GPU if available. As is
standard practice in non-linear optimization problems, we add a
Tikhonov regularization term to prevent the optimizer from getting
stuck in otherwise flat parts of the objective function.

%Several techniques are used to make the convergence fast and more robust:
%%
%\begin{enumerate}
%\item Use of a point-by-point goal for E and r\_E for the high frequency matrix
%\item Use the JAX library for automatic differentiation of Python and NumPy code to perform the gradient calculation needed by L-BFGS [13]
%\item Constraint matrix entries to be between –1 and 1
%\item Add a Tikhonov regularization term to the objective function
%\end{enumerate}%
%%

With large arrays, we find that using an existing matrix for the
initial guess (\vec{x_0}), for example an AllRAD design, ensures quick
convergence.  For smaller arrays, the initial guess can be a random
matrix. Running time for small arrays is a few seconds, larger arrays
can take a couple of minutes.

%% WARNING
%% we need to locate this figure here so it appears in the _next_ page
%% which is the one that talks about the stage, we might need to move
%% this around unless we find a better way to do this
%%
\begin{figure*}[t]
  \begin{center}
    \includegraphics[width=\textwidth]{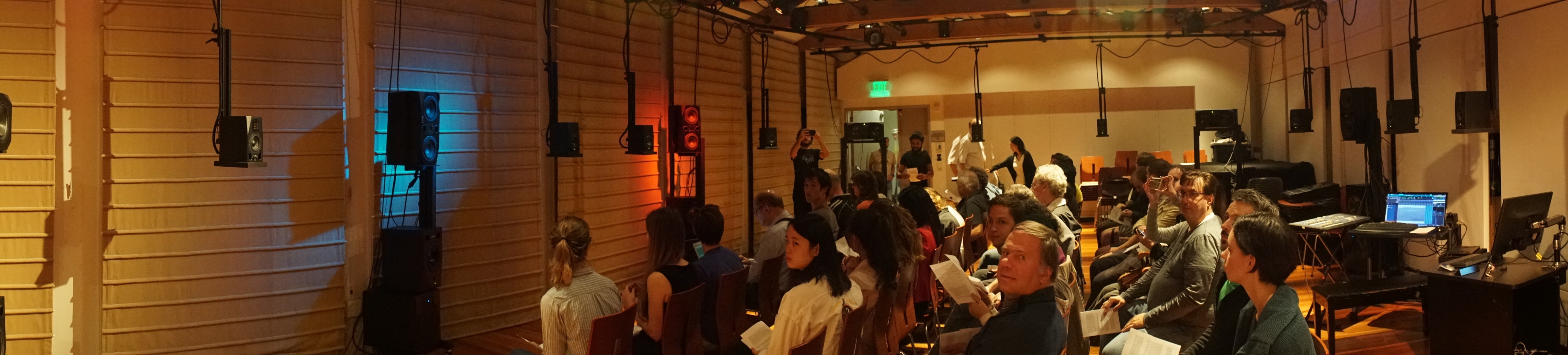}
    \caption{The Stage at CCRMA.  This is a permanent installation of 
    56 full-range loudspeakers.}
    \label{fg:stage_wide}
  \end{center}
  \vspace{-10pt}
\end{figure*}

\subsection{Optimizing \rEvec and \rVvec for Mixed-Order Signal Sets}
For full-order systems, there are closed-form expressions for the
maximum achievable magnitude of \rEvec for a particular Ambisonic
order such as Table 3.5 in \cite{Moreau:2006wr}, but none (that we
know of) exist for mixed-order signal sets.  Using too large a value
in the objective function makes the convergence behavior less
robust. Our solution is to design a mixed-order decoder for a
spherical-design 240-loudspeaker array with the desired mixed-order
signal set.  Due to the integration properties of a spherical design,
this is a well-behaved optimization problem and yields an optimal
decoder matrix.  We then compute \rEvec for this matrix at each point
in a 5200-point spherical design and use those values as the goal for each
corresponding direction in the optimization process for the actual
loudspeaker array.%

In a second step, the low-frequency matrix is optimized with the goal
that \rVvec points in the same direction as \rEvec,
$\uvec{\rE} = \uvec{\rV}$ and has a
magnitude of 1 over the area covered by the speaker array,
$|\rVvec| = 1$, thereby satisfying criteria (5) and (2), respectively.

\subsection{Tikhonov Regularization Problems}
Initial tests showed that the while the Tikhonov regularization term
sped up convergence, it also tended to shut off loudspeakers, using
the minimum needed for the signal set. While this may be desirable for
``spectral impairment'' considerations \cite{Solvang:2008p18}, we have
found that keeping those speakers active increases the size of the
sweet spot \cite{BLaH11}.  Another consideration is that some arrays,
such as the Stage array at CCRMA, use a mixture of speaker types,
where some speakers are full-range, while others have limited bass
response and power handling capability. We added an optional, per-speaker ``spareness penalty'' to
the objective function to allow a user to specify that some speakers
should not be turned off by the optimizer.

\begin{figure}[b!]
  \vspace{2pt}
  \begin{center}
    \includegraphics[width=0.91\columnwidth]{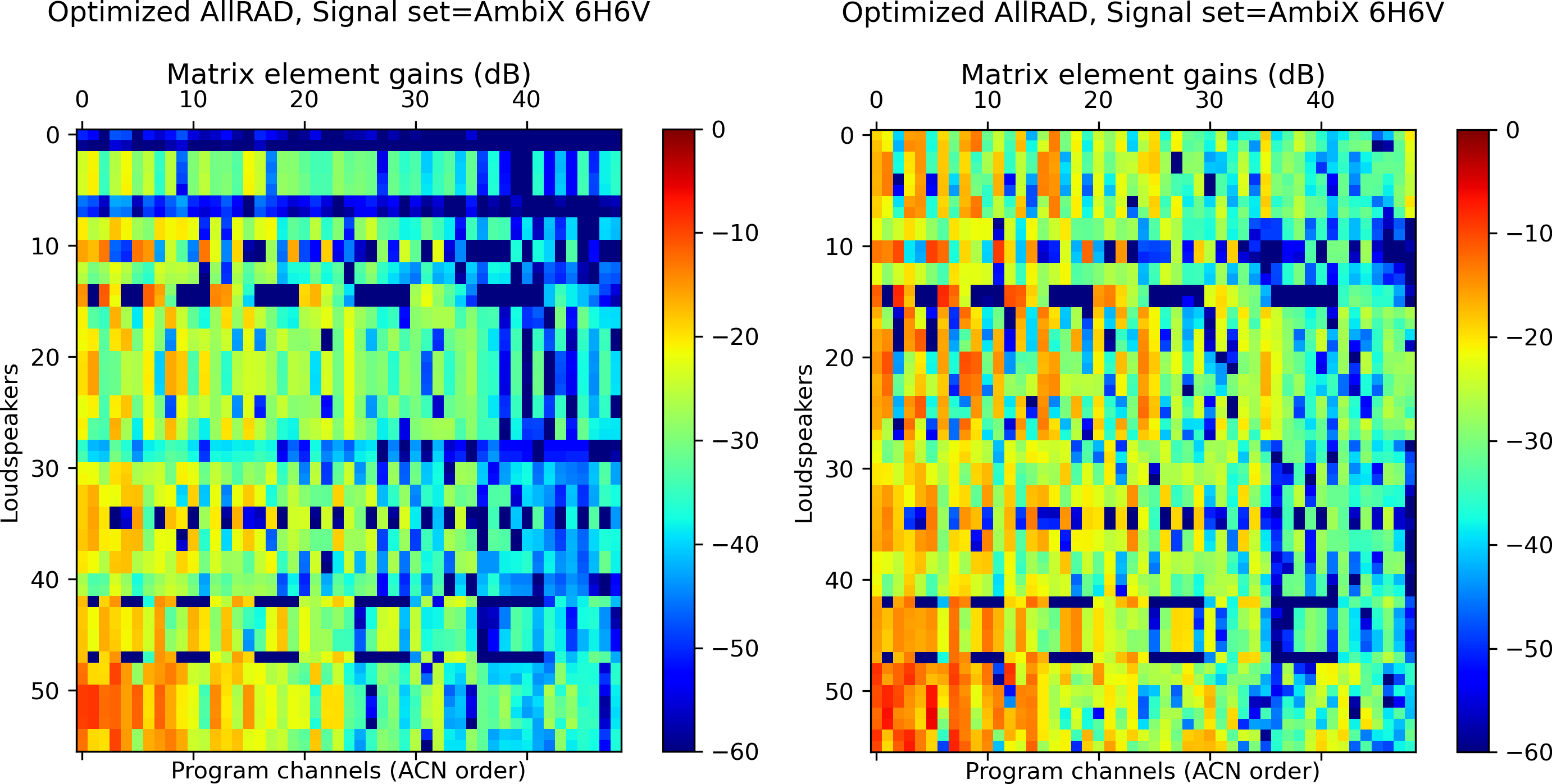}
    \caption{Stage 6th order matrix, sparseness\_penalty=0 in left plot, 1.0 in right plot}
    \label{fg:stage_matrix_sp}
  \end{center}
  \vspace{-20pt}
\end{figure}

The left pane of \figureref{fg:stage_matrix_sp} shows how speakers 1,
2, 7, and 8 in the Stage speaker array (four of the 8 big speakers in
the ear-level layer) are being turned off by the optimizer when the
sparseness penalty is 0. When setting sparse penalty to 1.0 (right
pane) those speakers are again active and contributing to the decoded
sound field.

\section{Results and Discussion}

We studied several loudspeaker arrays and found that in each case, the
optimizer produced equal to or better-performing decoders than
standard techniques such as AllRAD and EPAD. In the following
sections, we show examples based on a large and small array.

\subsection{The Stage at CCRMA}

The Stage, shown in \figureref{fg:stage_wide}, is a small concert
space at CCRMA, Stanford University.  It has a permanently-mounted
array of 56 full-range speakers and eight subwoofers. Of the twenty
speakers comprising the main ``ear-level'' ring, eight are larger and
located in movable tower stands. The ear-level ring and the upper
speakers form a fairly uniform dome of 48 speakers. There are eight
additional speakers at floor level, on the bottom of the eight main
tower stands, that were added to anchor the decoded sound image so
that sounds coming from the horizontal plane are not elevated. The
eight towers also house the subwoofers.

\begin{figure}[t!]
\centering
\begin{tabular}{c}
    \includegraphics[width=0.91\columnwidth]{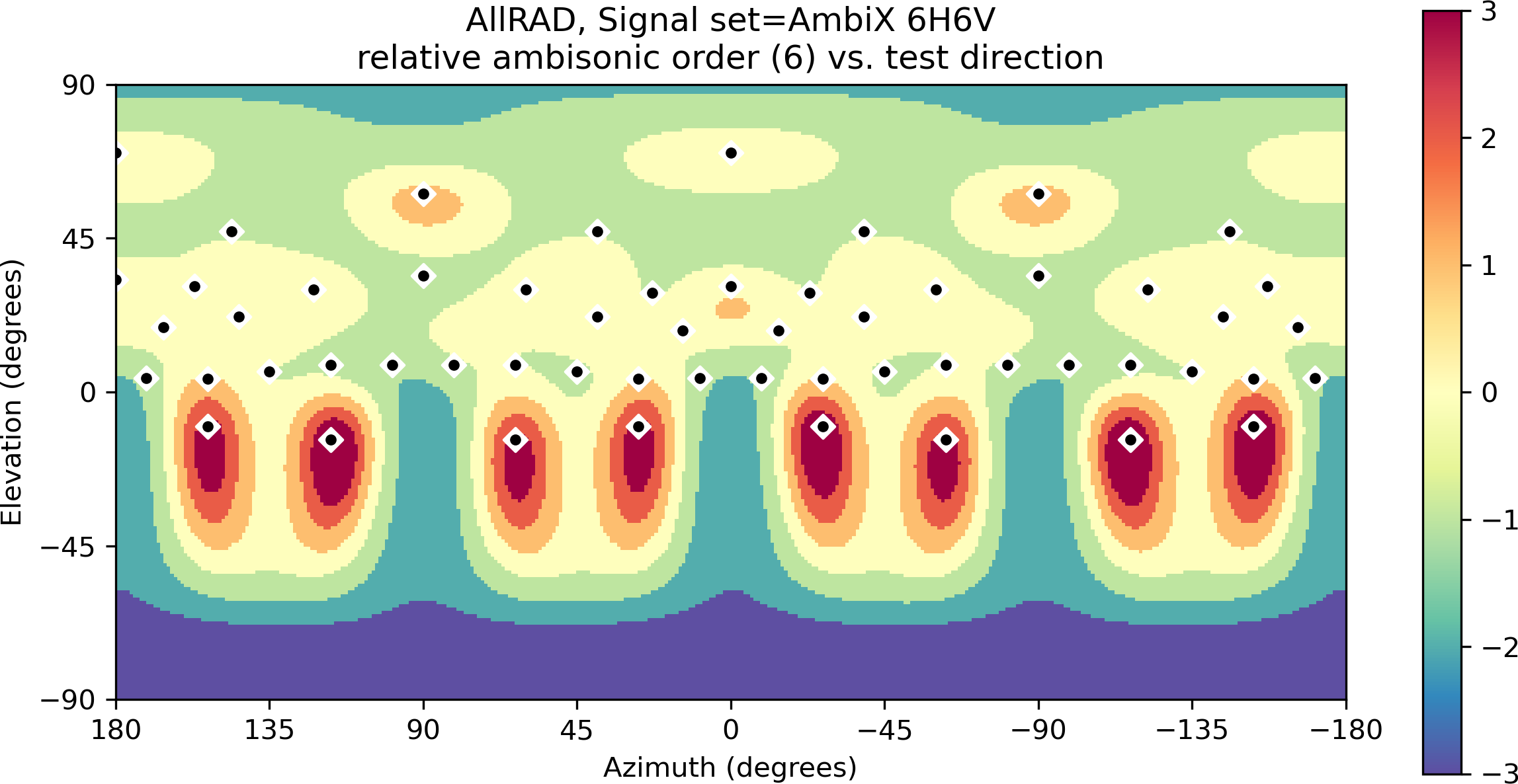} \\
    (a) 6H6V AllRad\\
    \includegraphics[width=0.91\columnwidth]{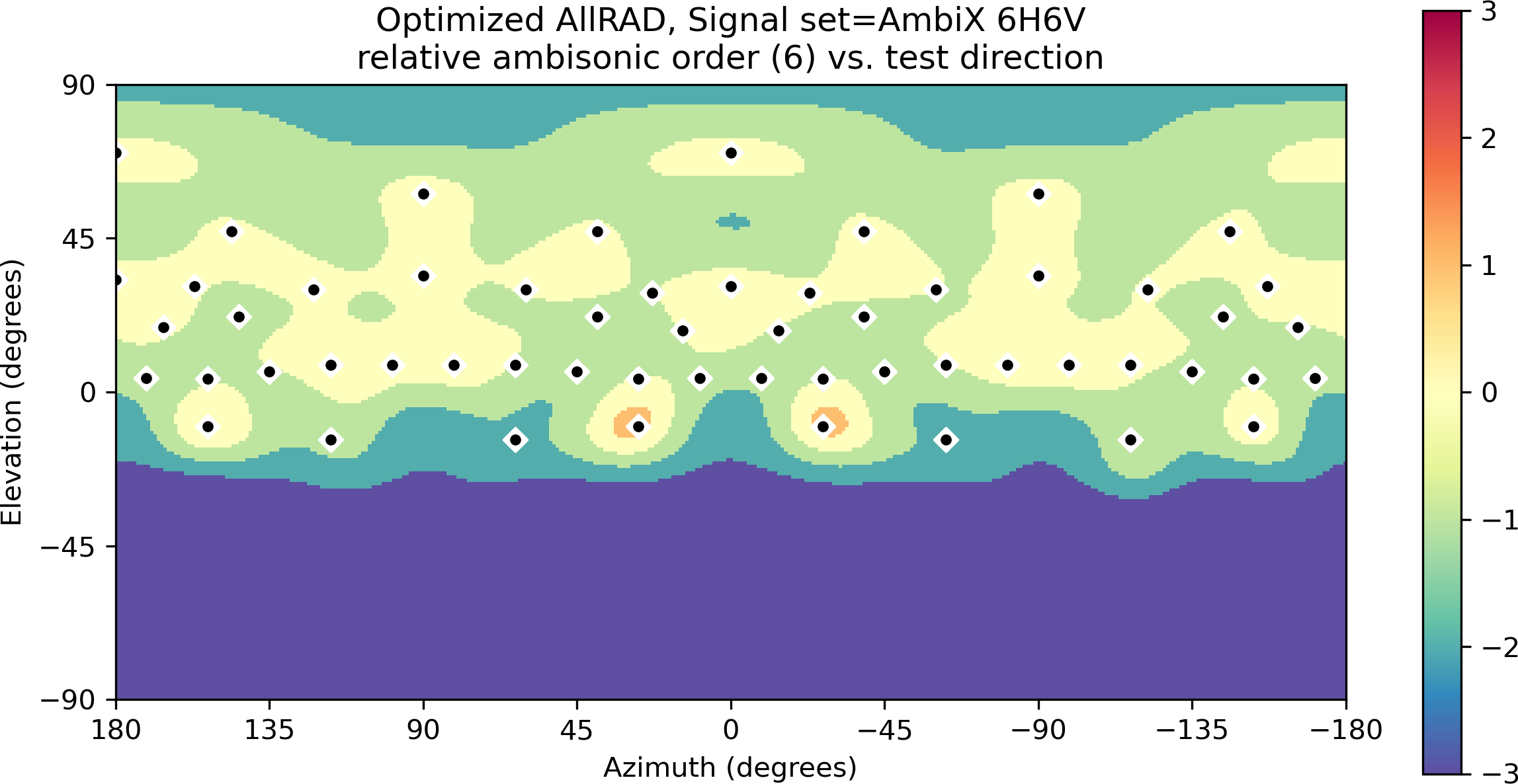} \\
    (b) 6H6V optimized, sparseness\_penalty=0.5\\ 
    \includegraphics[width=0.91\columnwidth]{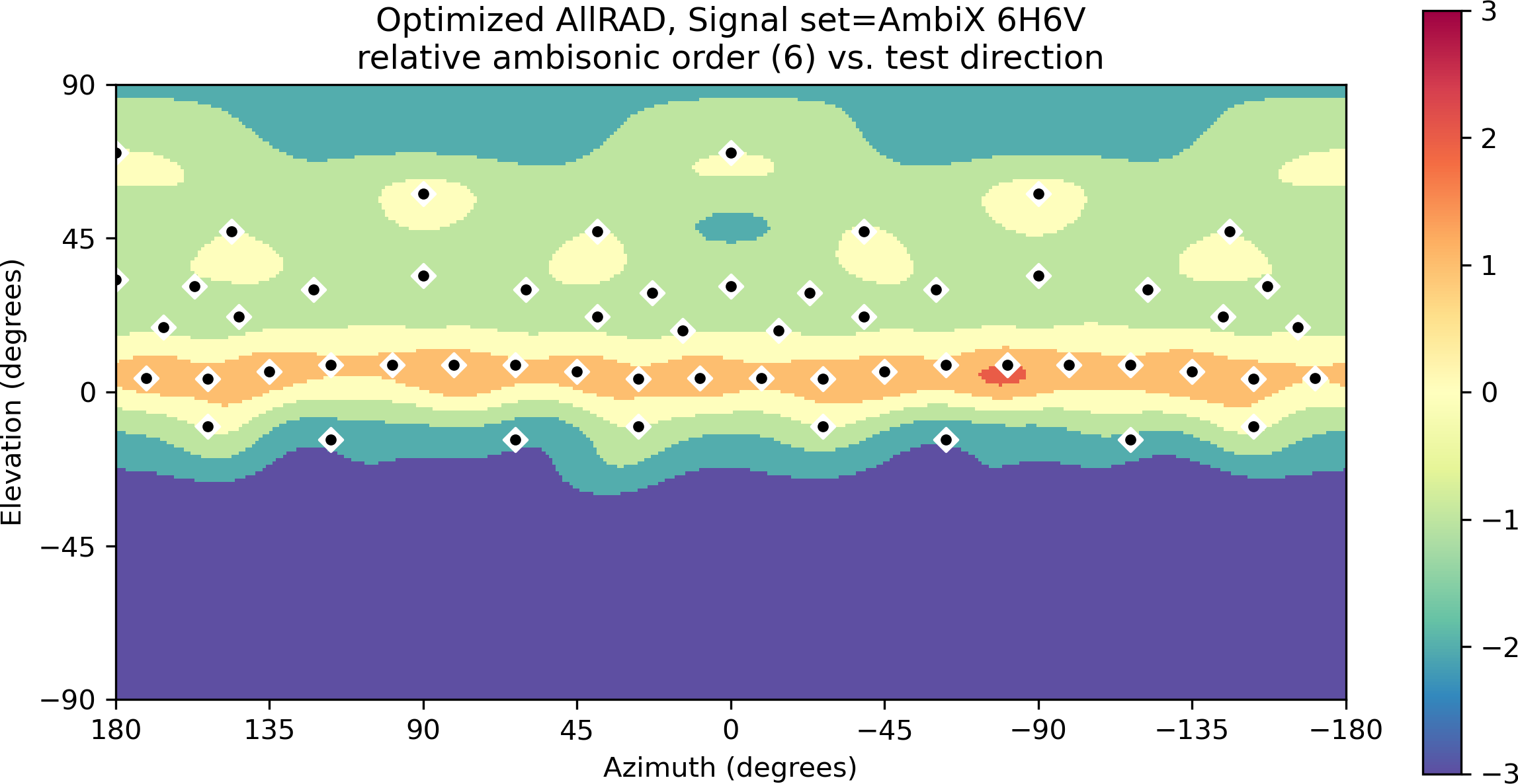} \\
    (c) 6H6V optimized, sparseness\_penalty=1.0\\\
\end{tabular}
\caption{Stage array, reproduction quality relative to sixth order.\textsuperscript{1}}
\label{fig:stage-sixth-order}
\end{figure}

As noted, the 48 upper speakers are distributed uniformly, but the eight
lower ones form a significantly sparser ring when compared to the rest
of the array, and are not evenly spaced in the horizontal plane. This
is a challenge to existing decoder design methods. Figure
\ref{fig:stage-sixth-order}(a) shows the relative Ambisonic order of a 6H6V
AllRAD decoder.%
\footnote{In these plots, the value of rE is calculated, then mapped to the nearest corresponding Ambisonic order and then displayed relative to the design order of the decoder.  Hence ``0'' indicates the decoder is operating at its designed order, "+1" is one up, "-1" one less and so forth, this shows how good the rendering quality is in different directions}

The performance in the region just below the horizon
which is rendered mainly by these eight speakers is very uneven. The areas
of high $|\rEvec|$ indicate that sound will tend to jump from one
speaker to the next as it is panned around the array and hence, they
do not assist in smoothing the performance in the vicinity of the horizontal plane.

\begin{figure}[t!]
\centering
\begin{tabular}{c}
    \includegraphics[width=0.91\columnwidth]{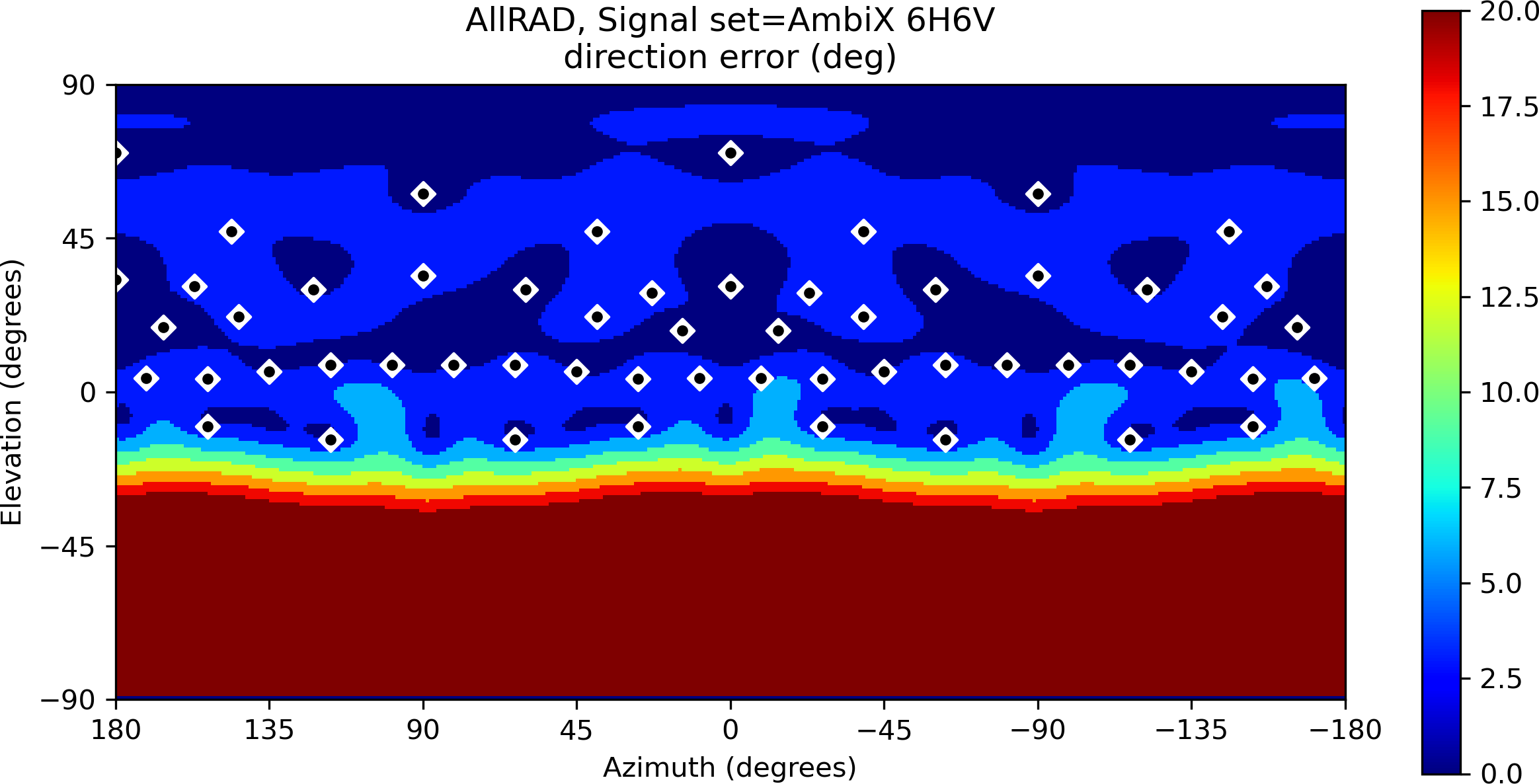} \\
    (a) 6H6V AllRad\\
    \includegraphics[width=0.91\columnwidth]{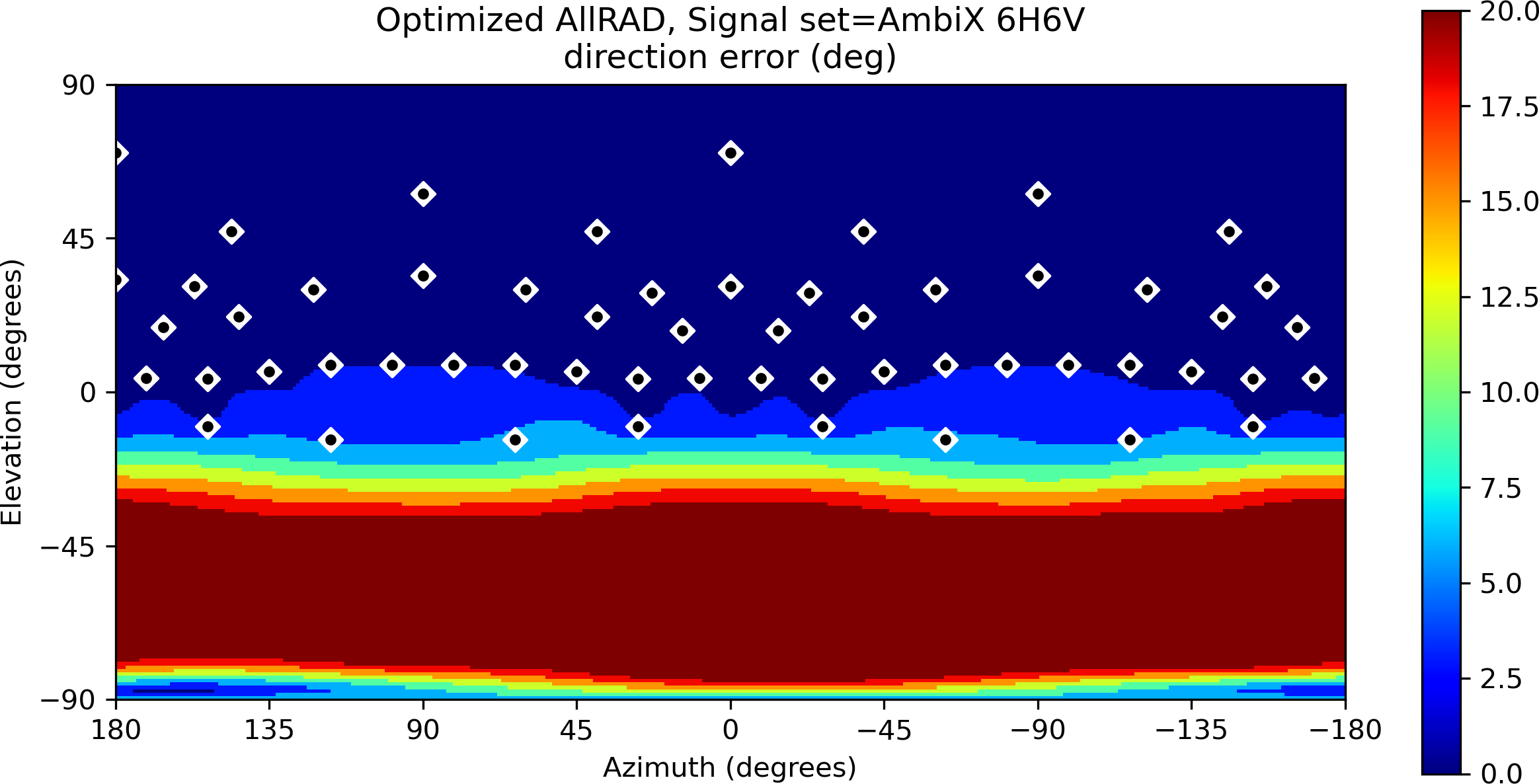} \\
    (b) 6H6V optimized\\
\end{tabular}
\caption{Stage array, direction error relative to intended direction (clipped at 20 degrees).}
\label{fig:stage-6H6V-de}
\end{figure}

%\begin{figure}[ht!]
%  \begin{center}
%    \includegraphics[width=1.0\columnwidth]{figures/stage_amb_order_crop.png}
%    \caption{Stage 6th order AllRAD reproduction quality relative to 6th order.}
%    \label{fg:stage_allrad}
%  \end{center}
%  \vspace{-10pt}
%\end{figure}

An optimized decoder (\figureref{fig:stage-sixth-order}(b),
sparseness\_penalty=0.5) trades off performance at low elevations for a
smooth integration of the lowest eight speakers into the full
array. Note how there is a wider vertical band around the main ring in
which the desired order of decoding happens correctly.

%\begin{figure}[ht!]
%  \begin{center}
%    \includegraphics[width=1.0\columnwidth]{figures/stage_amb_order_opt_crop.png}
%    \caption{Stage 6th order optimized decoder, sparseness penalty=0.5.}
%    \label{fg:stage_allrad_o}
%  \end{center}
%  \vspace{-10pt}
%\end{figure}

The sparseness\_penalty parameter can be used to exert some control over
how the speakers are used by the optimization process. For this array,
setting it to ``1'' enhances the horizontal performance of the array
at the cost of reduced performance at high elevations (Figure
\ref{fig:stage-sixth-order}(c), sparseness\_penalty=1).

%\begin{figure}[ht!]
%  \begin{center}
%    \includegraphics[width=1.0\columnwidth]{figures/stage_amb_order_opt_sp1_crop.png}
%    \caption{Stage 6th order optimized relative order, sparseness penalty=1.0.}
%    \label{fg:stage_allrad_o_sp1}
%  \end{center}
%  \vspace{-10pt}
%\end{figure}

\figureref{fig:stage-6H6V-de} shows the
direction error for both the (a) AllRAD and (b) optimized decoder. In this
case the improvement is marginal as the original decoder is already
performing very well.

%\begin{figure}[ht!]
%  \begin{center}
%    \includegraphics[width=1.0\columnwidth]{figures/stage_dir_error_spkrs_crop.png}
%    \caption{Stage 6th order AllRAD decoder, direction error of \rEvec relative to intended direction.}
%    \label{fg:stage_allrad_de}
%  \end{center}
%  \vspace{-10pt}
%\end{figure}
%
%\begin{figure}[ht!]
%  \begin{center}
%    \includegraphics[width=1.0\columnwidth]{figures/stage_dir_error_spkrs_opt_crop.png}
%    \caption{Stage 6th order optimized decoder, direction error of \rEvec relative to intended direction.}
%    \label{fg:stage_opt_de}
%  \end{center}
%  \vspace{-10pt}
%\end{figure}

\subsection{Home Dome 8+5 speaker array}

\begin{figure}[b!]
  \begin{center}
    \includegraphics[width=0.91\columnwidth]{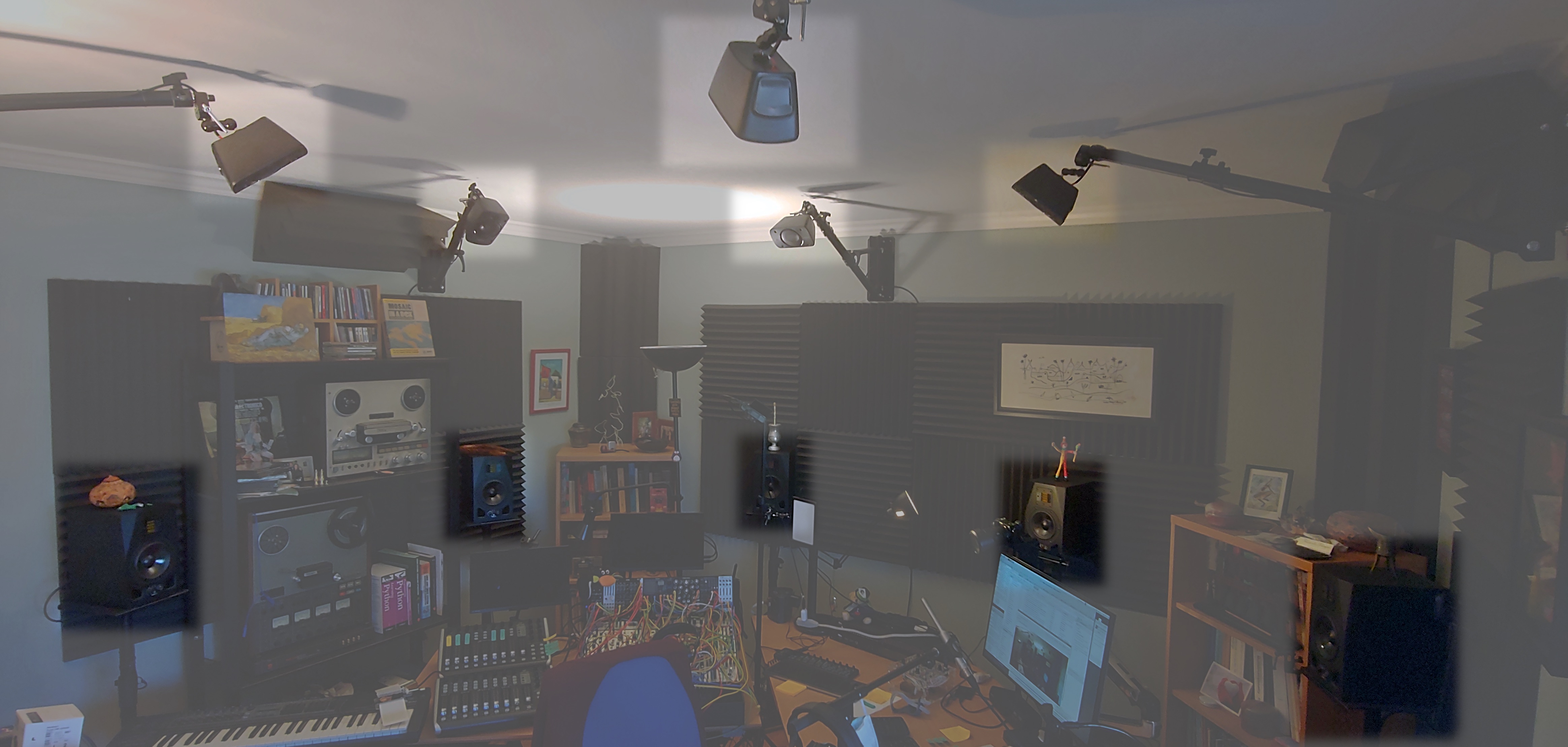}
    \caption{Home Dome, eight horizontal and five height speakers. Speaker positions highlighted.}
    \label{fg:home_studio}
  \end{center}
  \vspace{-10pt}
\end{figure}

This is a small, two-ring speaker array with an ear-level ring of
eight speakers and an upper ring of five smaller speakers at 45 degree
elevation. It is a good example of an array that needs an optimized
mixed-order decoder.

The lower-eight speaker ring can render third-order horizontal, but
the combined array can only do mixed-order rendering.

%% Most decoders are designed for a full set of Ambisonics components,
%% but they perform poorly when fed a mixed-order signal set. < insert
%% more detailed explanation that removing components (making them
%% zero) is not the same as designing a decoder that does not take
%% into account those components > The following figures [[ STILL
%% MISSING ]] illustrate the directional errors that occur when a 3H3V
%% decoder is fed with 3H1V or 3H2V signal sets.

\begin{figure}[b!]
\centering
\begin{tabular}{c}
    \includegraphics[width=0.91\columnwidth]{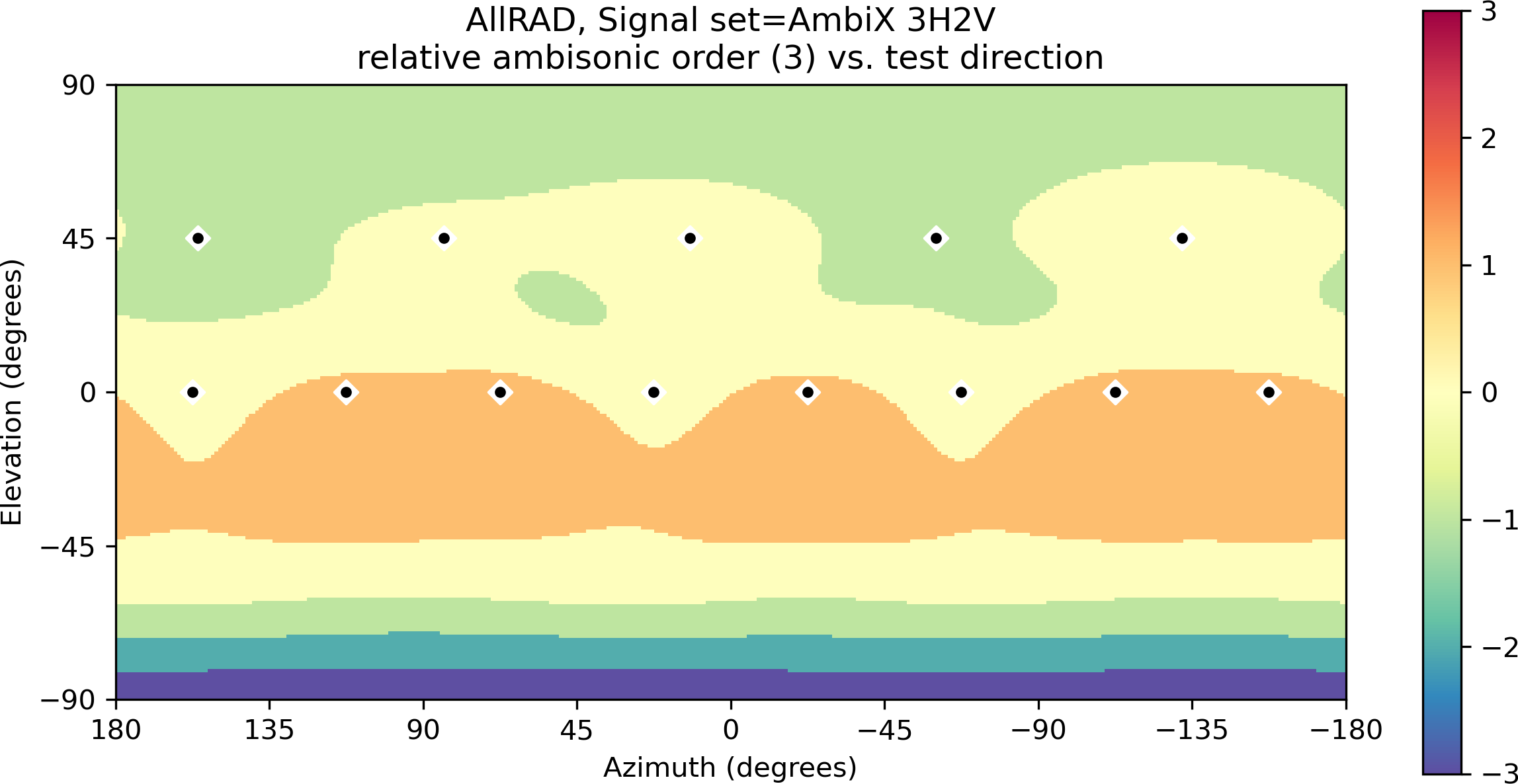} \\
    (a) 3H2V AllRad\\
    \includegraphics[width=0.91\columnwidth]{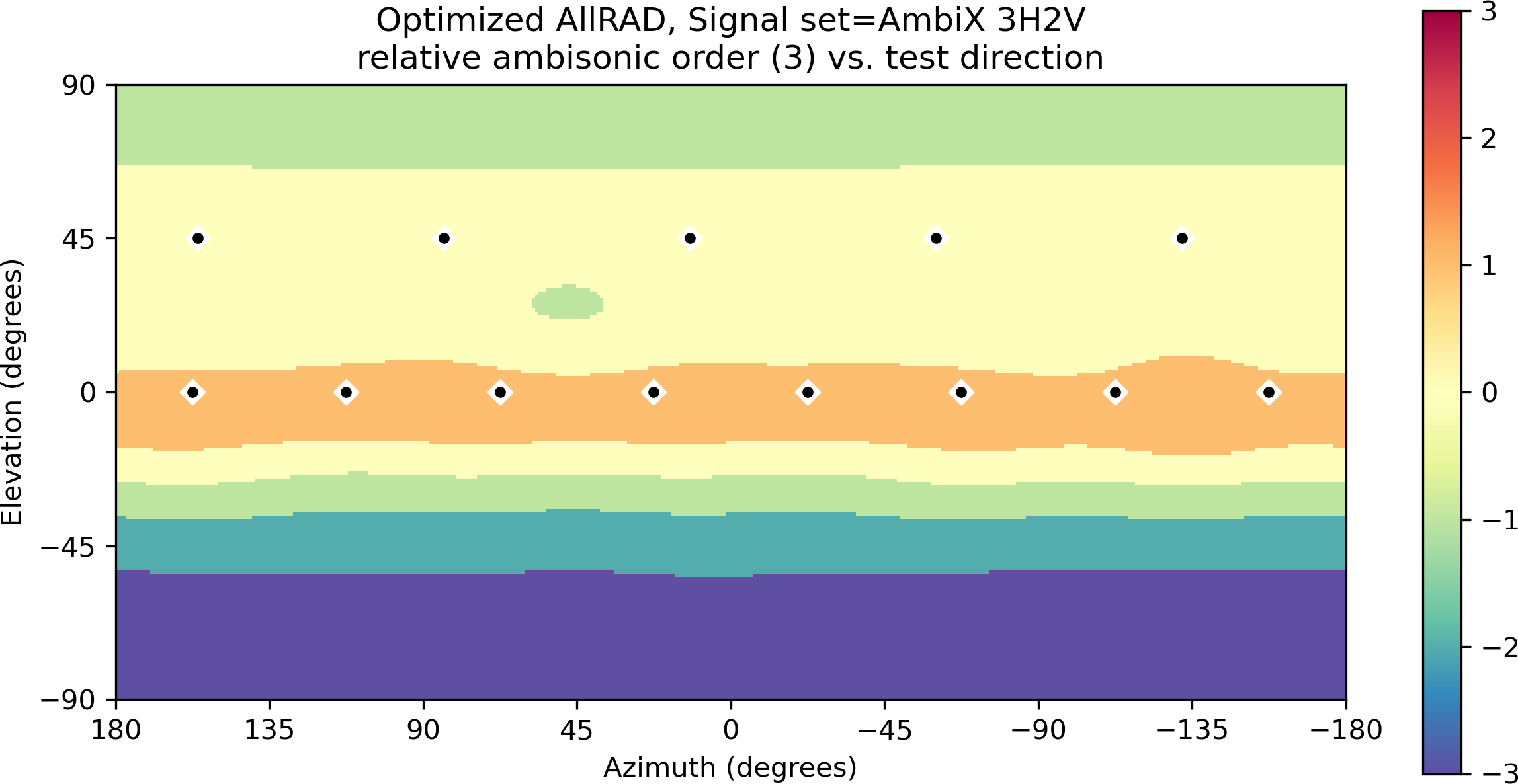} \\
    (b) 3H2V optimized\\
\end{tabular}
\caption{Home Dome decoders, reproduction quality relative to third order.}
\label{fig:nando-3H2V}
\end{figure}

\figureref{fig:nando-3H2V}(a) shows the relative order performance of
an AllRAD 3H2V mixed-order decoder for this array. An optimized
mixed-order decoder, \figureref{fig:nando-3H2V}(b), creates a much more
even rendering of the sound field, with higher performance in the
horizontal plane and even performance in the dome above the
listener. Performance suffers at the top of the dome, as is to be
expected because of the lower speaker density there.

%\begin{figure}[ht!]
%\begin{center}
%\includegraphics[width=1.0\columnwidth]{figures/nando_allrad_amb_order_crop.png}
%\caption{Home Dome, AllRAD Decoder, reproduction quality relative to third order.}
%\label{fg:nando_allrad}
%\end{center}
%\vspace{-10pt}
%\end{figure}
%
%\begin{figure}[ht!]
%\begin{center}
%\includegraphics[width=1.0\columnwidth]{figures/nando_allrad_amb_order_opt_crop.png}
%\caption{Home Dome, optimized decoder, reproduction quality relative to third order.}
%\label{fg:nando_opt}
%\end{center}
%\vspace{-10pt}
%\end{figure}

\begin{figure}[b!]
\centering
\begin{tabular}{c}
    \includegraphics[width=0.91\columnwidth]{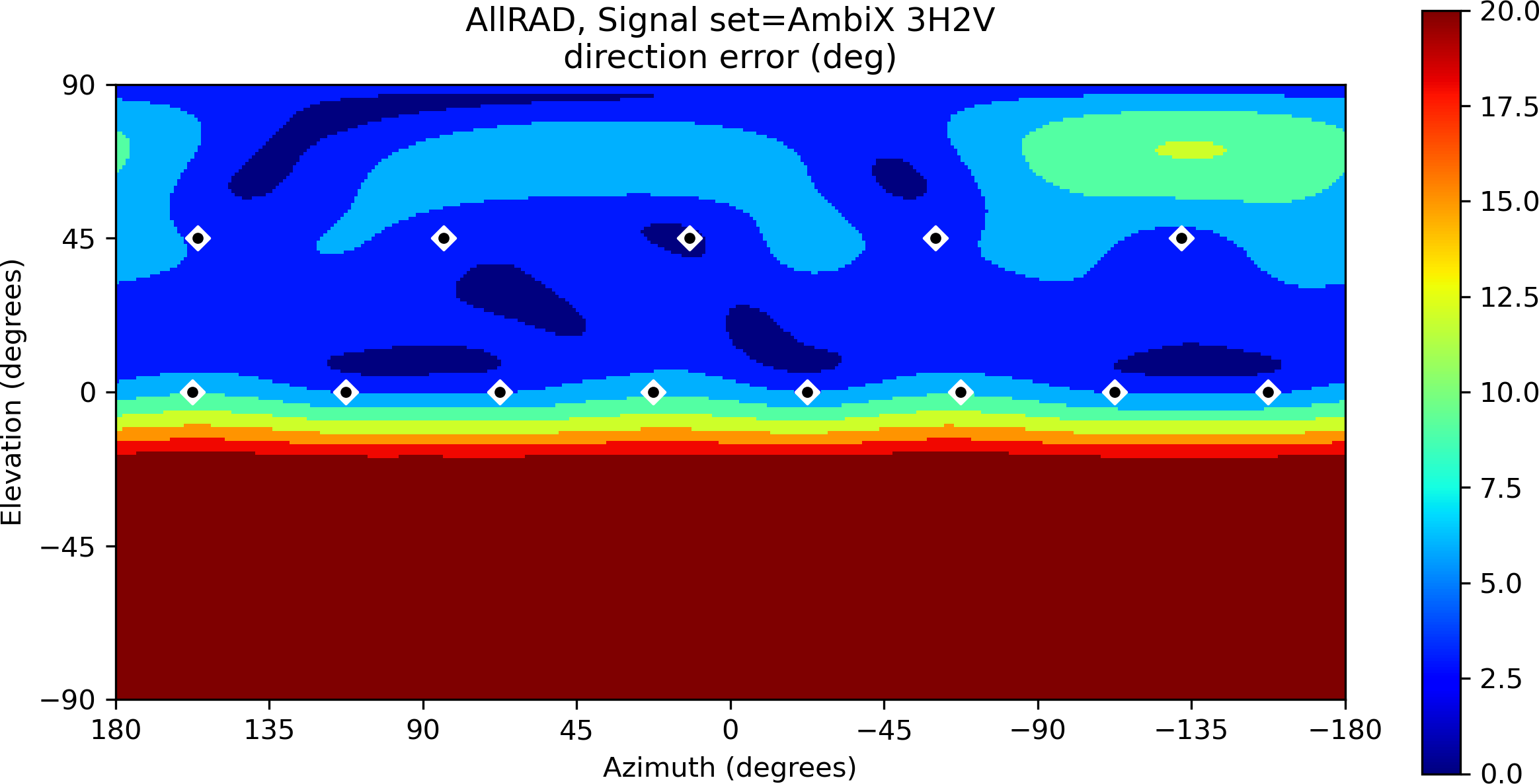} \\
    (a) 3H2V AllRad\\
    \includegraphics[width=0.91\columnwidth]{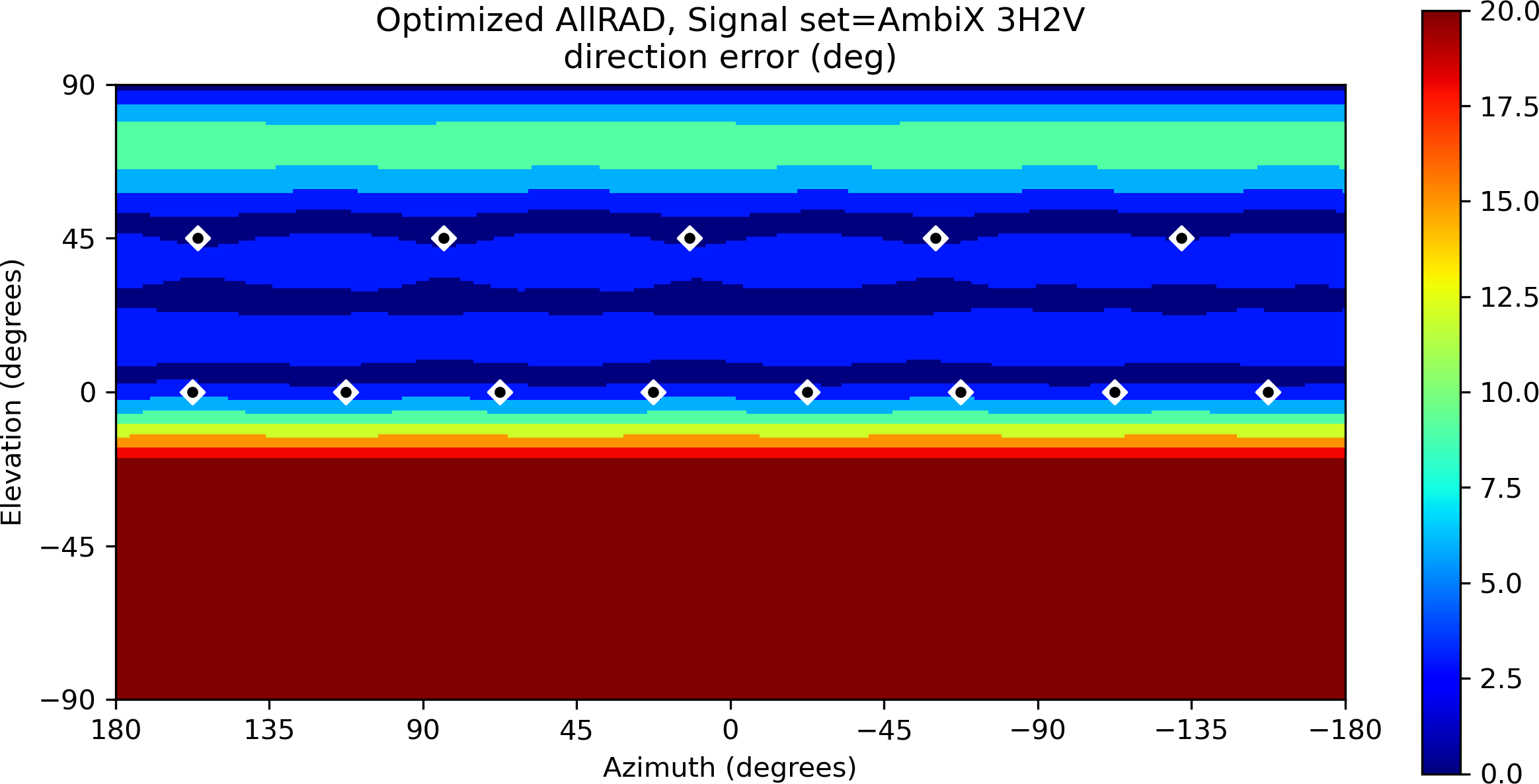} \\
    (b) 3H2V optimized\\
\end{tabular}
\caption{Home Dome decoders, direction error relative to intended direction (clipped at 20 degrees).}
\label{fig:nando-3H2V-de}
\end{figure}

Direction error is also minimized by the optimized decoder (\figureref{fig:nando-3H2V-de}). The improvement is significant in this small array.

%
%\begin{figure}[ht!]
%\begin{center}
%\includegraphics[width=1.0\columnwidth]{figures/nando_allrad_dir_error_spkrs_crop.png}
%\caption{Home Dome AllRAD decoder, direction error of \rEvec relative to intended direction.}
%\label{fg:nando_allrad_de}
%\end{center}
%\vspace{-10pt}
%\end{figure}
%
%\begin{figure}[ht!]
%\begin{center}
%\includegraphics[width=1.0\columnwidth]{figures/nando_allrad_dir_error_spkrs_opt_crop.png}
%\caption{Home Dome optimized decoder, direction error of \rEvec relative to intended direction.}
%\label{fg:nando_opt_de}
%\end{center}
%\vspace{-10pt}
%\end{figure}

\begin{figure}[t!]
\centering
\begin{tabular}{c}
    \includegraphics[width=0.91\columnwidth]{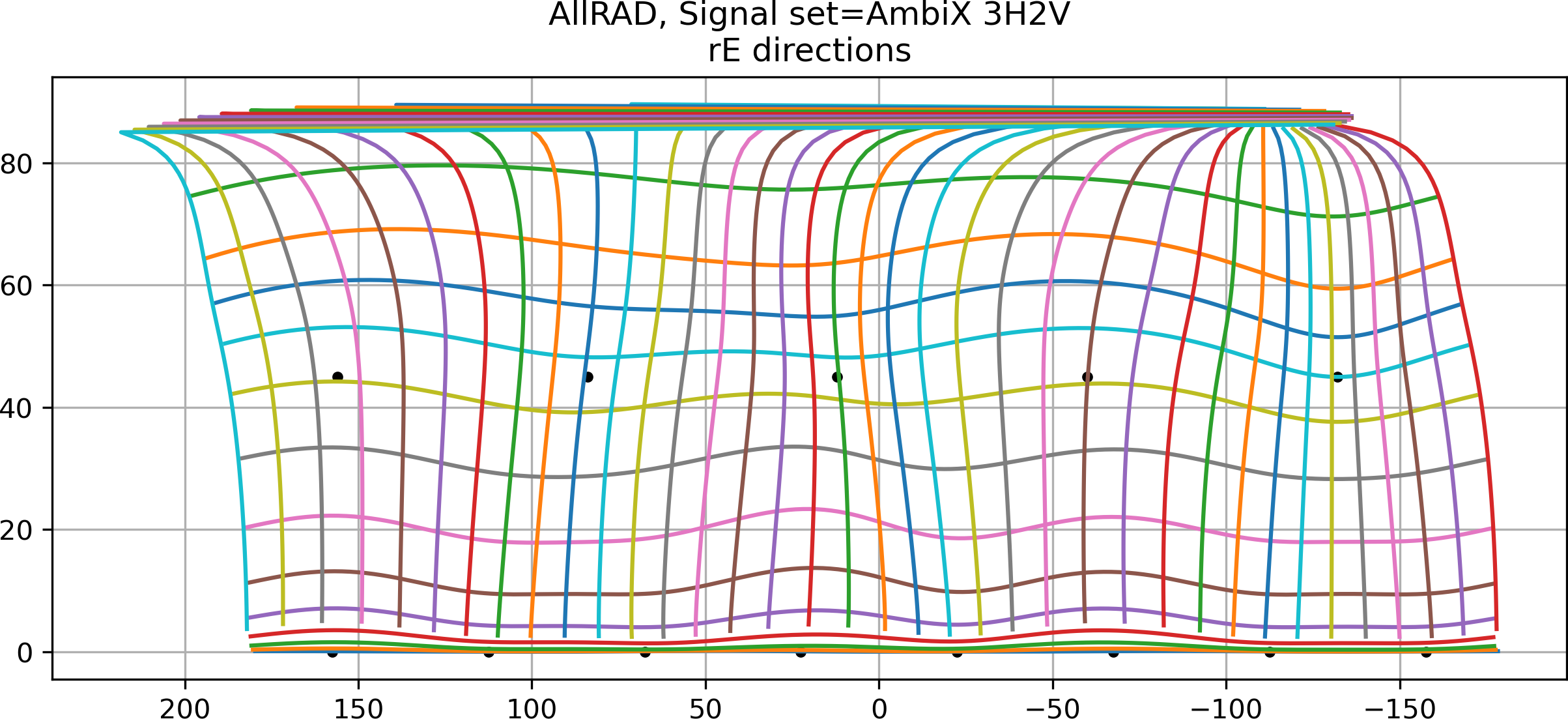} \\
    (a) 3H2V AllRad\\
    \includegraphics[width=0.91\columnwidth]{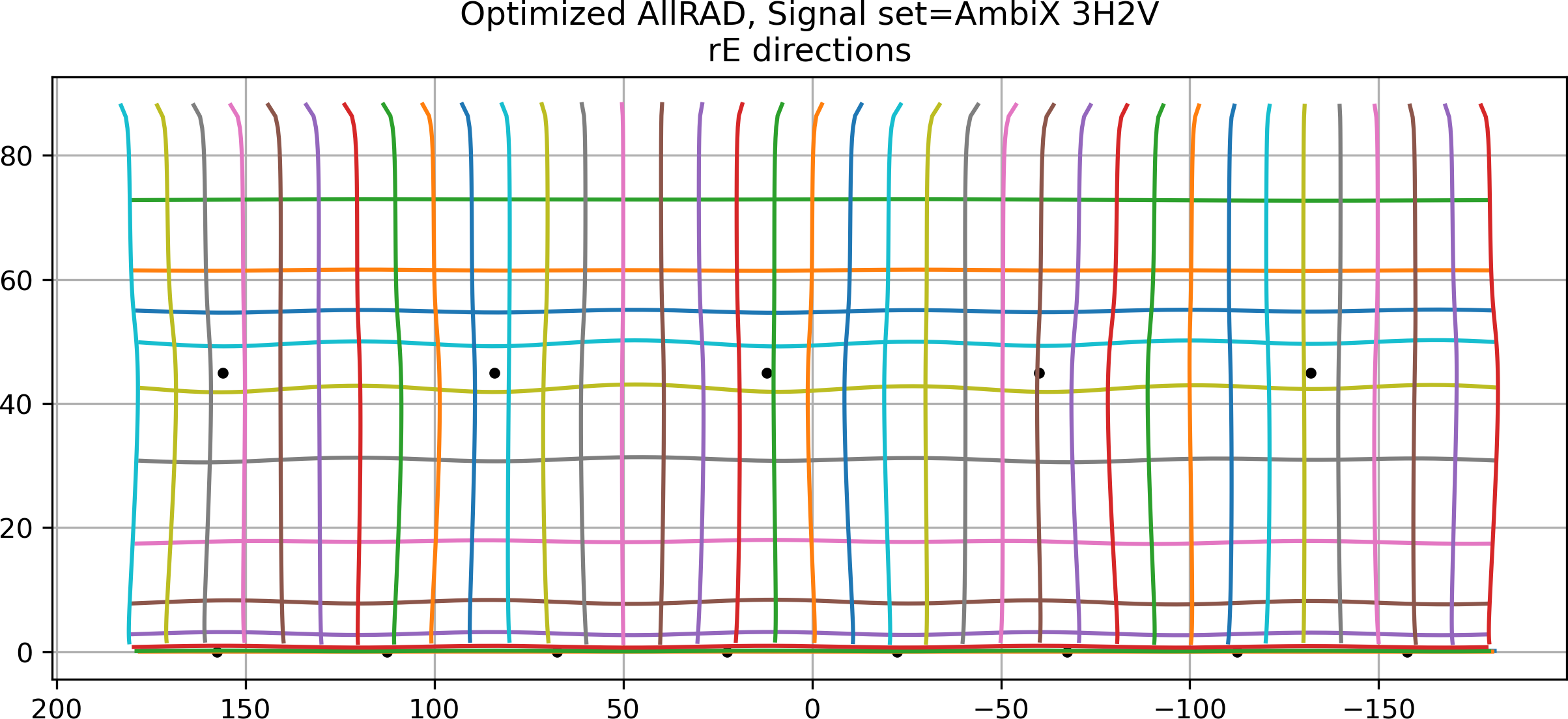} \\
    (b) 3H2V optimized\\
\end{tabular}
\caption{Home Dome decoders, rendered position of source moving along lines of constant azimuth and elevation.}
\label{fig:nando-3H2V-grid}
\end{figure}

Another way of looking at the directional performance is to plot the
actual directions from which sources would originate if moving along
lines of constant azimuth or elevation. We can see these plots in \figureref{fig:nando-3H2V-grid} which show that the optimized decoder has much less directional error than the plain
AllRAD.

%\begin{figure}[ht!]
%\begin{center}
%\includegraphics[width=1.0\columnwidth]{figures/nando_allrad_dirs_crop.png}
%\caption{Home Dome AllRAD decoder, rendered position of source moving along lines of constant azimuth and elevation.}
%\label{fg:nando_allrad_d}
%\end{center}
%\vspace{-10pt}
%\end{figure}
%
%\begin{figure}[ht!]
%\begin{center}
%\includegraphics[width=1.0\columnwidth]{figures/nando_allrad_dirs_opt_crop.png}
%\caption{Home Dome optimized decoder, rendered position of source moving along lines of constant azimuth and elevation.}
%\label{fg:nando_opt_d}
%\end{center}
%\vspace{-10pt}
%\end{figure}

\begin{figure}[t!]
\centering
\begin{tabular}{c}
    \includegraphics[width=0.91\columnwidth]{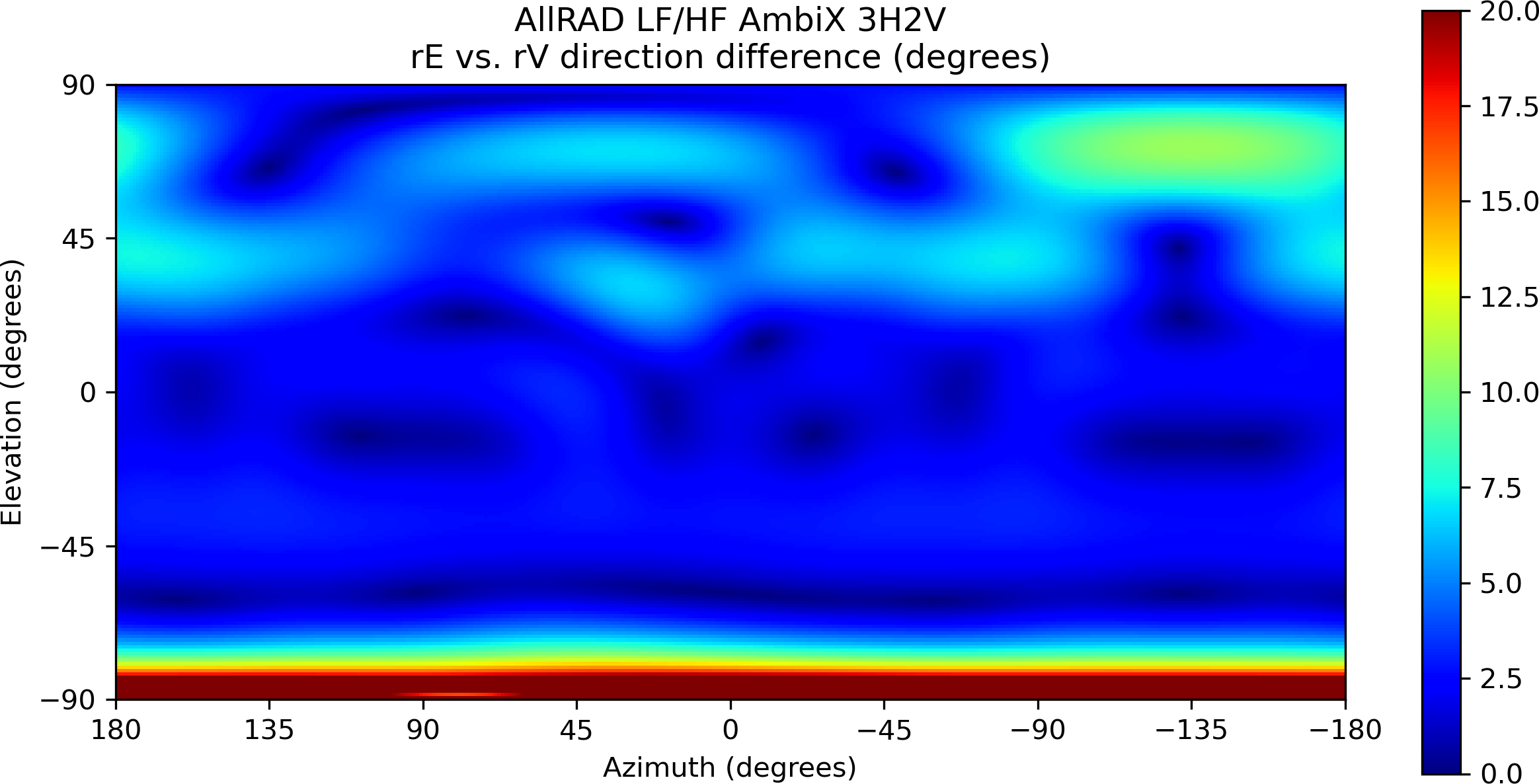} \\
    (a) 3H2V AllRad\\
    \includegraphics[width=0.91\columnwidth]{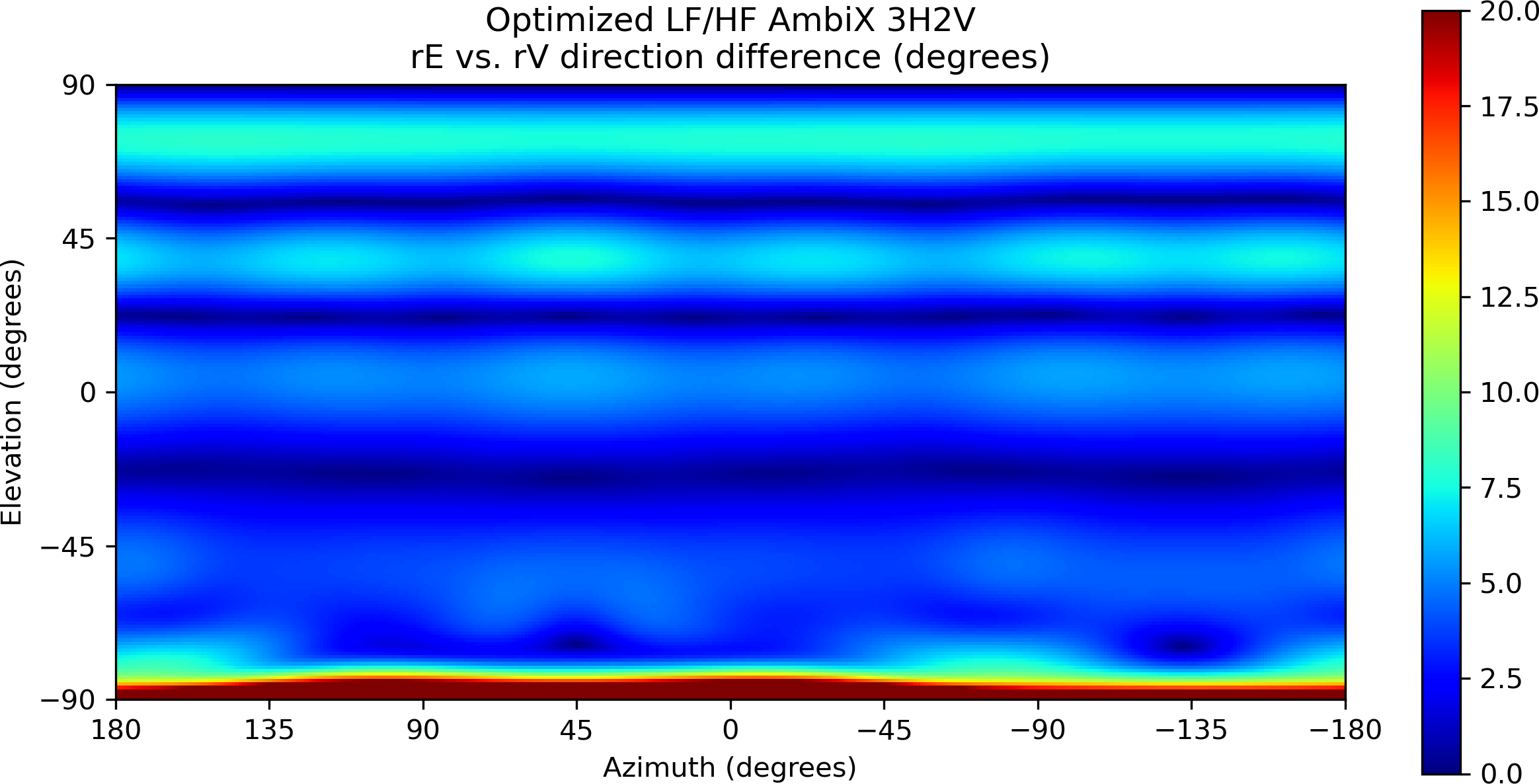} \\
    (b) 3H2V optimized\\
\end{tabular}
\caption{Home Dome decoder, \rEvec vs. \rVvec direction error.}
\label{fig:nando-3H2V-rvre_d}
\end{figure}

\figureref{fig:nando-3H2V-rvre_d}
shows how adding the second optimization stage for \rVvec and using a
two-matrix ``Vienna'' style decoder minimizes the directional mismatch between low- and
high-frequency performance.   In addition to
pointing in the correct direction, the magnitude of \rVvec has made uniformly 0.95-1.0 (the ideal value), from a starting point that varied between 0.5 and 1.5 depending on direction.

%\begin{figure}[ht!]
%\begin{center}
%\includegraphics[width=1.0\columnwidth]{figures/nando_allrad_rv_dir_error_crop.png}
%\caption{Home Dome AllRAD \rEvec vs. \rVvec direction error.}
%\label{fg:nando_allrad_rvre_d}
%\end{center}
%\vspace{-10pt}
%\end{figure}
%%
%\begin{figure}[ht!]
%\begin{center}
%\includegraphics[width=1.0\columnwidth]{figures/nando_opt_rv_dir_error_crop.png}
%\caption{Home Dome optimized \rEvec vs. \rVvec direction error.}
%\label{fg:nando_opt_rvre_d}
%\end{center}
%\vspace{-10pt}
%\end{figure}

%\section{Discussion}
%
%We have conducted preliminary informal listening tests on domestic
%arrays with promising results. We plan to do more as COVID-19
%restrictions at Stanford and SRI are lifted and will report in future
%papers.

\section{Summary}
We describe a new implementation of the ADT
which has tools that more perfectly and quickly optimize the array
performance for the Ambisonic criteria. These tools are applied to the
design of two loudspeaker arrays, a 56-loudspeaker professional
installation, and a 13-loudspeaker array in a domestic
installation. The analysis tools were applied to the question of
whether it is acceptable to use a full-order decoder with a mixed-order signal set. Analysis shows that in every case it is better to
derive a separate mixed-order decoder.

Early work on Ambisonics described decoders for either 2D or 3D
regular arrays of loudspeakers. Most applications in the real world
involve arrays that are either irregular or incomplete. One example is
approximate hemispherical arrays. Such arrays are inherently irregular
and the missing bottom half makes the resulting decoder have
increasing error at and below the horizon.  There is a second problem
having to do with mixed-order decoders. It is frequently the case that
the density of loudspeakers is less for directions above the
horizontal. This happens either because of the expense of the
additional loudspeakers or because of difficulties mounting the speakers. In this
case, the array has a different capability in different directions,
almost always with greater performance for the horizontal direction
than for height. Also, some microphone arrays have different Ambisonic order
performance in horizontal and vertical directions.

Subsequent work described methods such as AllRAD for deriving decoders
for these arrays. These methods result in decoders that only
approximately meet the Ambisonic criteria. Numerical optimization
methods can be used to enable the decoder to have nearly perfect
behavior through sparse regions and at the edge of array coverage.  By
``nearly perfect'' we mean as good as possible for the number of
loudspeakers available. The visualization tools provided with the ADT
enable the designer to make choices as to which ambisonic parameters
are to be optimized. These can be used, for instance, to determine how
many of a given total number of loudspeakers are to be used for
horizontal and how many for height.

Informal listening tests were performed for the two loudspeaker arrays
described above but were limited to only a single listener due to the
COVID-19 restrictions at Stanford and SRI. Future work will
include more formal tests with a larger set of listeners.

The code described in this paper is free and open source, and can be
accessed via the ADT repository\cite{www:adt}. The implementation is a work in
progress, but is fully capable of producing working (and very good)
decoders. The code can be run via Jupyter notebooks using Google
Research's Colaboratory facility \cite{www:colab}. We provide notebooks
that reproduce the results in this paper.  The FAUST code produced
can be compiled online as well, producing plugins for most types of 
audio processing programs \cite{www:faust-editor}.

\bibliographystyle{jaes}
\renewcommand*{\bibfont}{\small}
\raggedright
% Reference to bibliography file.
\bibliography{ambisonics, IFC-18, refs}

\end{document}